\documentclass[twocolumn,showpacs,preprintnumbers,amsmath,amssymb]{revtex4}

\usepackage{graphicx}
\usepackage{dcolumn}
\usepackage{bm}
\usepackage{natbib}

\begin{document}

\title{Phase sensitive Brillouin scattering measurements with a novel magneto-optic modulator}

\author{F. Fohr}
\author{A.A. Serga}%
 \author{T. Schneider}%
\author{J. Hamrle}%
\author{B. Hillebrands}%

\affiliation{%
Fachbereich Physik and Forschungszentrum OPTIMAS, Technische Universit\"at 
Kaiserslautern, Erwin-Schr\"odinger-Stra\ss e 56, D-67663 Kaiserslautern, Germany
}%

\date{\today}
\begin{abstract}
A recently reported phase sensitive Brillouin light scattering technique is improved by use of a magnetic modulator. This modulator is based on Brillouin light scattering in a thin ferrite film. Using this magnetic modulator in time- and space Brillouin light scattering measurements we have increased phase contrast and excluded influence of optical inhomogeneities in the sample. We also demonstrate that the quality of the resulting interference patterns can be improved by data postprocessing using the simultaneously recorded information about the reference light.
\end{abstract}

\maketitle

\section{Introduction}
Space- and time-resolved Brillouin light scattering (BLS) spectroscopy is a well established technique to
investigate the spin-wave dynamics in thin magnetic films \cite{Demokritov} . However, this method is based on a simple
counting of inelastically scattered photons. Thus it only allows for the spatial and temporal mapping of
spin-wave intensities. No phase information about magnetic excitations is accessible by conventional BLS.
At the same time the phase information is crucial to answer questions concerning problems such as formation of
coherent states in a magnon gas and evolution of nonlinear spin-wave eigenmodes \cite{Moebius} as well
as nonlinear phase accumulation \cite{Schneider1}, peculiarities of spin-wave excitation process
\cite{Schneider2}, 2-dimensional phase structure of spin-wave beams in magnetically anisotropic media,
etc.\\
\begin{figure}
\includegraphics[width=0.5\textwidth]{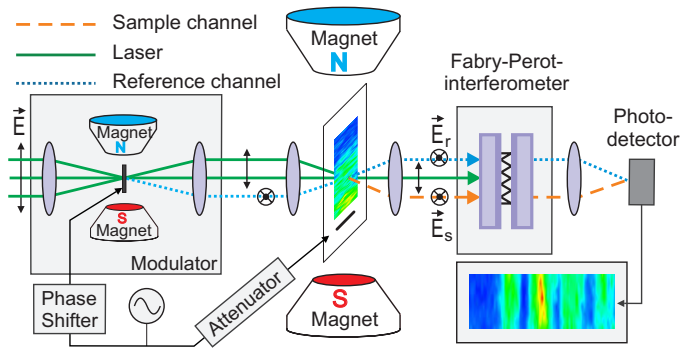}
\caption{(Color online) Experimental setup: phase resolved BLS. Phase  sensitivity is created by interference between the light inelastically scattered by the spin waves (dashed line) and a coherent reference beam (dotted line), frequency shifted and turned in polarization by the magneto-optic modulator.}
\label{figure1} 
\end{figure} 

The first realization of a phase-sensitive BLS setup as well as the results obtained by means of
phase-sensitive BLS spectroscopy have been presented by us in Ref.~\cite{Schneider1, Schneider2, Serga}.
It was shown that the implementation of phase resolution into Brillouin light spectroscopy leads to a
complete picture of the underlying physical processes by combining space-, time- and phase-resolution
into the measurement. \\
Here we report on further improvement of this technique by implementation of a new type of magneto-optic
modulator, based on Brillouin light scattering in a thin ferrite film. Furthermore we demonstrate that the use of the
magneto-optic modulator also improves the BLS dynamical range.

\section{Principle of operation}
To understand the principle of phase sensitivity it is important to notice that the inelastically
scattered light, which is detected in the BLS measurements, is determined by both amplitude and phase
of the scattering spin wave. However, the phase information is lost when the photon is received by
a photodetector, it only can register the arrival of a photon, i.e. quantum of energy. \\
To access phase information we use interference between inelastically scattered light created in the process of propagation of the probing laser beam through the sample with reference scattered light provided by the magneto-optic modulation. \\Without a fixed phase correlation, the intensity of the sample beam $E^2_\mathrm{s}$ and the reference beam 
$E^2_\mathrm{r}$ combine to 
\begin{eqnarray}
E^2_\mathrm{s}+E^2_\mathrm{r}
\label{eq:one}.
\end{eqnarray}
If conditions for interference are fulfilled, the sample beam and the reference beam combine to
\begin{eqnarray}
E^2_\mathrm{s}+2E_\mathrm{s}E_\mathrm{r} \cos(\varphi)+E^2_\mathrm{r}
\label{eq:two}.
\end{eqnarray}
with an additional phase-term.\\

The magneto-optic modulator used here was produced on the base of an yttrium-iron-garnet (YIG) ferrite
film. As one can see in Fig.~\ref{figure1} the laser beam is initially focused on a 10 $\mu m$ thick
in-plane magnetized YIG film stripe in the magneto-optic modulator. The bias magnetic field is oriented
parallel to the spin-wave propagation direction, and thus a backward volume magnetostatic spin wave
(BVMSW) \cite{Damon} is excited in the film. As a result of nonelastic light scattering by this wave, a part of the
beam is now frequency shifted and simultaneously rotated in polarization by 90$^\circ$ forming the
reference beam. Focussing of the laser beam is necessary to enhance the efficiency of the scattering
process. Firstly, the beam is focussed near to the exciting antenna, where the intensity of the spin wave
and scattering-probability is highest and secondly, as we have a propagating and not a standing spin wave, the phase of the scattered light is well defined only if the  scattering process occurs in an area which length is small compared to the wavelength of the spin wave.\\
Note that a great part of the laser light passes the modulator unchanged in frequency and polarization.
Some part of this undisturbed light undergoes a second scattering process in the sample and forms the
sample beam. Having the same polarization direction and the same frequency, the reference beam and the sample
beam now fulfill the condition for interference at the photodetector.\\
The modulator is driven by the same microwave signal which is used to excite the spin waves. This guarantees not only
that the reference light has exactly the same frequency as the inelastically scattered light but also
the necessary phase coherence between both signals. The amplitude of the reference signal and the phase difference $\varphi$ can be
adjusted  by using a microwave attenuator and a phase shifter respectively (see Fig.~\ref{figure1}). Since the light in the signal and the reference
channnels share the same spatial path, thermal and mechanical stability is ensured.\\
The resulting interference picture allows us to visualize the phase fronts and to calculate the phase
profiles (i.e.\ the time-dependent phase difference between the exciting microwave signal and the spin
wave at any given point) of the investigated spin waves from measured interference maps with different
additional phase shifts between reference and signal. For a more detailed description of the underlying
analysis procedure and the phase-sensitive BLS setup see Refs.~\cite{Schneider1, Schneider2}.\\
Before the implementation of the {\it magneto}-optic modulator, the necessary frequency shift for the
reference beam was created by using an {\it electro}-optic modulator based on a lithium-niobate
crystal \cite{Serga,Caponi}. However phase-resolved BLS spectroscopy of magnetic excitations using the proposed magneto-optic modulation has
several advantages compared to the previous electro-optic-modulation.\\
{\it  (i)} In order to increase the conversion efficiency, the lithium-niobate-crystal in our electro-optic
modulator was placed inside a microwave cavity. Due to the strong distortion of the electric field
inside the cavity by influence of this dielectric ($\varepsilon =40$) material, the adjustment
of the modulator was very complicated and as a result only a few fixed frequencies were accessible. This
restriction does not exist for a magneto-optic modulator which can be tuned continuously in frequency by an
external magnetic field. Thus, the RF-frequency can be implemented as an additional free parameter
during a measurement.

{\it  (ii)} Another important advantage of a magneto-optic modulator, which is based on Brillouin light
scattering, is that the polarization of the reference beam and the sample beam is the same, i.e. both beams 
change polarization with respect to the light source. On contrary, the electro-optic modulator does not change the polarization of the frequency-shifted light
and as a result no interference with non-elastically scattered light occurs without special disalignment
of the polarization filter placed before the interferometer. In conventional BLS spectroscopy this
filter is used to block the elastically-scattered light and thus consequently decreases the 
noise. In this way the disalignment of the filter, necessary when using an electro-optic modulator,
allows interference but also significantly increases the noise level. This disalignment is not necessary for the proposed magneto-optic modulator, where the
polarization condition for the best interference is fulfilled automatically.

\section{Experimental results}

\begin{figure}
\includegraphics[width=0.4\textwidth]{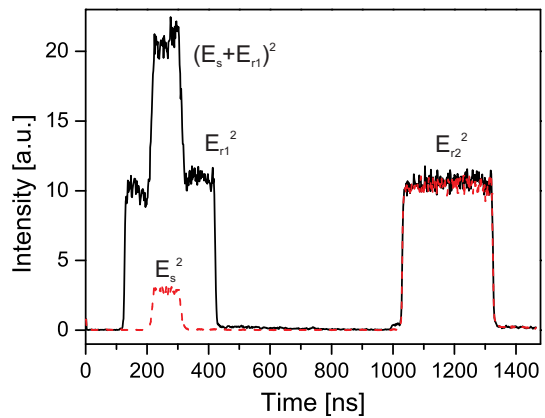}
\caption{(Color online) Time profile example of the interference between sample pulse $E^2_\mathrm{s}$ and reference pulse $E^2_\mathrm{r1}$. An additional reference pulse $E^2_\mathrm{r2}$ is recorded with a time delay of 900 ns.}
\label{figure2} 
\end{figure} 

\begin{figure}
\includegraphics[width=0.45\textwidth]{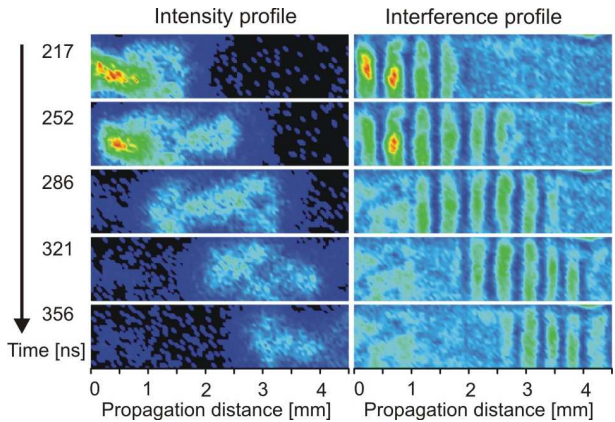}
\caption{(Color online) Phase resolved measurement of a propagating spin-wave packet using a magneto-optic modulator. The first column is obtained with the conventional space- and time-resolved Brillouin light scattering setup and shows two-dimensional maps of the spin-wave intensity distribution for given times. The right column shows the corresponding interference patterns. Each profile is normalized to the respective maximum value at 217 ns.}
\label{figure3} 
\end{figure}

In the following we present example measurements with the new setup. Spin-wave pulses are excited in a 4,5 mm long YIG-film in BVMSW-configuration. 
Figure~\ref{figure2} shows the comparison between the time profiles of the sample pulse ($E^2_\mathrm{s}$, dashed line) and the constructive interference ($(E_\mathrm{s}+E_\mathrm{r1})^2$, solid line) between sample pulse and reference pulse $E^2_\mathrm{r1}$. The data for the intensity of the sample pulse is obtained by switching off the exciting microwave-current in the modulator. The duration is 300 ns for the reference pulse and 100 ns for the sample pulse, respectively. The pulse carrier frequency is 7 GHz at a bias magnetic field $H_0$ of 1815 Oe. To obtain the information about spatial and temporal distribution of the background, an additional reference pulse $E^2_\mathrm{r2}$ was recorded with a time delay of 900 ns. \\
The spatial distribution of the propagating spin-wave packet is shown in Fig.~\ref{figure3}. The first column shows the decaying intensity of the spin-wave packet propagating in the sample.  In the second column the interference pattern of the sample-beam and reference beam is depicted. It is clearly observable that, for the long propagation times, the spin wave is much more visible in the interference patterns in comparison to the intensity profiles. Thus an increase in the dynamic range of the BLS setup is obtained.\\
\begin{figure}
\includegraphics[width=0.4\textwidth]{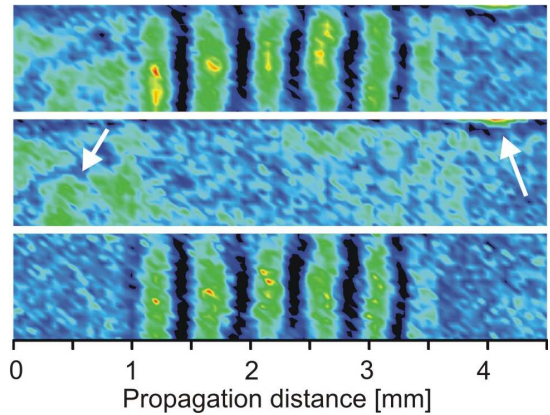}
\caption{(Color online) Phase resolved measurement of a spin-wave packet using a magneto-optic modulator. The first two panels show the intensity map of the interference pattern, described by Eq.~(\ref{eq:two}) at 286 ns and the time delayed reference pulse $E^2_\mathrm{r2}$. In the third panel the interference pattern is divided by the intensity profile of the time delayed reference beam, resulting in Eq.~(\ref{eq:three}). Thus signal variations caused by optic nonuniformities are removed (white arrows). Each profile is normalized to the respective maximum value at 286 ns.}
\label{figure4} 
\end{figure}
The sample beam in Eq.~(\ref{eq:two}) contains all information about the spin wave and the pure reference component $E^2_\mathrm{r}$  can be regarded as a background, including defects and optic nonuniformities in the YIG-film. The resulting picture can be improved by just dividing the interference profile by the intensity profile of the time delayed reference beam.
\begin{eqnarray}
\frac{E^2_\mathrm{s}}{E^2_\mathrm{r}} +2\frac{E_\mathrm{s}}{E_\mathrm{r}} \cos(\varphi)+1
\label{eq:three}.
\end{eqnarray}
Since reference beam and sample beam share the same spatial path, both are influenced by the same disturbances. Thus by dividing both signals, the noise can be significantly reduced. The result of this calculation is presented in Fig.~\ref{figure4}. Spurious signals, e.g.\ the spot at the upper right corner, are completely erased from the interference pattern and the distribution of the spin wave itself can be distinguished better from the homogenous white noise.

In conclusion we have improved phase sensitive Brillouin light scattering spectroscopy by implementation of a new kind of magneto-optic modulator. \\

\begin{acknowledgments}
Support by the DFG (SFB/TRR 49, JST-DFG Hi380/21-1 and Graduiertenkolleg 792) is gratefully acknowleged.
\end{acknowledgments}

\end{document}